\documentclass[twocolumn, english, superscriptaddress, preprintnumbers, amsmath, amssymb, floatfix, aps]{revtex4-2}

\usepackage[T1]{fontenc}
\usepackage[latin9]{inputenc}
\usepackage{dcolumn}
\usepackage{bm}
\usepackage{graphicx}
\usepackage{color}
\usepackage{esint}
\usepackage{babel}
\usepackage{bbold}
\usepackage{amsfonts}
\usepackage{amsmath}
\usepackage{slashed}
\usepackage{bm}
\usepackage{epsfig}
\usepackage{enumerate}
\usepackage{blkarray}
\usepackage{booktabs}
\usepackage{multirow}
\usepackage[colorlinks=true, allcolors=blue]{hyperref}

\def\be{\begin{equation}}
\def\ee{\end{equation}}
\def\bea{\begin{eqnarray}}
\def\eea{\end{eqnarray}}
\def\bmat{\begin{pmatrix}}
\def\emat{\end{pmatrix}}
\def\bs{\begin{split}}
\def\es{\end{split}}

\def\~{$\approx$}
\def\bra{\langle}
\def\ket{\rangle}

\def\dag{\dagger}

\begin{document}

\title{Transport and spectral properties of magic angle twisted bilayer graphene junctions based on local orbital models}

\author{M. Alvarado}
\affiliation{Departamento de F{\'i}sica Te{\'o}rica de la Materia Condensada C-V, Condensed Matter Physics Center (IFIMAC) and Instituto Nicol\'as  Cabrera,  Universidad Aut{\'o}noma de Madrid, E-28049 Madrid, Spain} 

\author{A. Levy Yeyati}
\affiliation{Departamento de F{\'i}sica Te{\'o}rica de la Materia Condensada C-V, Condensed Matter Physics Center (IFIMAC) and Instituto Nicol\'as  Cabrera,  Universidad Aut{\'o}noma de Madrid, E-28049 Madrid, Spain} 

\date{\today}
\begin{abstract}
The electronic properties of junctions defined electrostatically on twisted bilayer graphene can be addressed theoretically using lattice models. Recent works 
have introduced minimal local orbital models to describe twisted bilayer graphene at the magic angle (MATBLG) with different degrees of approximation and accounting for fragile topology. In the present work we use a Green's function formalism to obtain the spectral and transport properties for MATBLG junctions based on these models. We introduce different symmetry breaking perturbations to simulate the effect of interactions and characterize the topology of the bulk bands by analyzing the corresponding Wilson loops. We then analyze the spectral properties for different types of edges and in the case of a domain wall in the sublattice symmetry breaking parameter. We further consider a three region junction where one could control independently the central and lateral regions doping level. In the limit where the central region is fixed at the charge neutrality point and the lateral ones are heavily doped, the spectral and the two terminal transport properties can be understood in terms of the hybridization of chiral states along the junctions. These properties are found to be extremely sensitive to the orientation of the junctions along the moir\'e lattice.      
\end{abstract}

\maketitle

\section{Introduction}

The ability to tune the electronic properties of Van der Waals 
heterostructures of two-dimensional materials by changing their relative angle has given rise to the new field of "twistronics" \cite{Carr2017,Ribeiro-Palau2018}. Twisted bilayer graphene is by now the most prominent example of this field, exhibiting a large variety of interesting phenomena. Near the so-called magic angle the flat bands close to charge neutrality arising from the large scale moir\'e pattern become extremely sensitive to electronic correlations and interactions, giving rise to a rich phase diagram including superconducting and insulating phases as a function of the doping level \cite{Cao2018,Cao2018_b}. 

The flat bands also exhibit topological properties \cite{Xia2020} which manifest for example in the appearance of anomalous quantum Hall effect \cite{Serlin2020, Andrei2021, Efetov2020}. More recently, the possibility to produce tunable junctions on this material by local electrostatic control has been demonstrated \cite{Rodan-Legrain2020} thus opening the way towards its application in transport devices.

From the point of view of theory, large efforts have been devoted to understand the origin of correlation induced phenomena in MATBLG \cite{Guinea2018, Guinea2020, Dai2021, Wilhelm2020, Zaletel2020, Yazdani2020, Stauber2018, Vafek2019, Yang2020, Carr2019}. A basic debate has emerged regarding the appropriate microscopic modeling. While the continuous model of Bistritzer and MacDonald \cite{Bistritzer2011} is broadly accepted as an accurate starting single-particle description, real space or lattice models could be more appropriate for the
inclusion of interactions or for the study of transport in MATBLG junctions. Lattice models suffer, however, of the so-called Wannier obstructions
\cite{thouless1984,soluyanov2011wannier, Balents2019,Senthil2018, Watanabe2018, Yang2019} 
which prevent the description of topological bands in terms of local orbitals. While 
simple low energy two bands models were proposed which avoid this problem by neglecting certain symmetries \cite{Koshino2018}, so called "faithful" tight-binding (TB) models including higher energy bands are able to account properly for
the fragile topology \cite{Wang2019, Bernevig2020, Wieder2018, BlancodePaz2019, Kooi2019, Li2020, Peri2020} of the low energy bands \cite{Vishwanath2019, Senthil2018}. Other models have been proposed mixing real and
momentum space descriptions \cite{Balents2021}, or including both valleys in a two-band description \cite{Yao2020}.

\begin{figure}[t]
\includegraphics[width=\columnwidth]{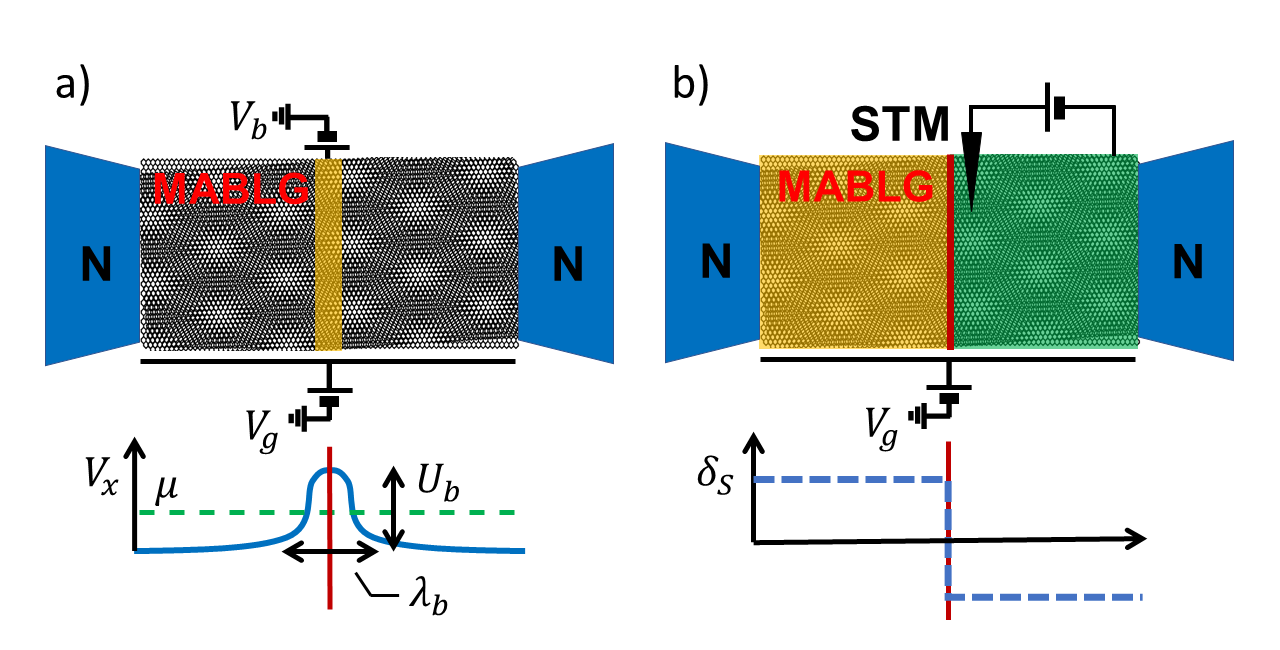}
\caption{Different experimental setups based on MATBLG analyzed in this work. a) Two terminal transport through a barrier defined by a backgate. The length of the barrier region can vary between $\lambda_b \approx 300-50$ nm. b) STM spectroscopy of a Domain Wall due to the interaction of MATBLG with the substrate which is incorporated in the model through the parameter $\delta_S$ (see text for details). The large size of the moir\'e pattern ($>10$ nm) in MATBLG would allow to study the effect of the alignement of the junction or the domain wall along well defined directions on the moir\'e lattice.}
\label{fig1}
\end{figure}

The aim of the present work is to analyze spectral and transport properties using lattice models for MATBLG junctions. We would like to address non-homogeneous situations like the ones depicted in Fig.~\ref{fig1} where either local spectral properties could be studied using STM techniques or two terminal transport measurements could be performed. Inspired by the recent experiments of Ref.~\cite{Rodan-Legrain2020} we study transport through regions with varying doping levels which could be tuned by gates as schematically shown in Fig.~\ref{fig1}~a). Additionally, our approach should allow us to predict the local spectral properties at such junctions or at the edges of TBLG samples, which could be accessed experimentally as demonstrated in recent STM studies on NbSe2/CrBr3 moir\'e superlattices \cite{Kezilebieke2020}. An important aspect of MATBLG which guides our study is that the large size of its moir\'e pattern ($>10$ nm) would allow transport experiments on junctions along well defined directions on the moir\'e lattice. As we show in this work, the junction's orientation plays an important role on its transport and spectral properties. Eventually our work should serve as a basis for analyzing transport and spectral properties in situations including superconducting phases. 

Our starting point are the lattice models from Refs \cite{Koshino2018} and \cite{Vishwanath2019} for which we construct the boundary Green functions (bGFs) using an efficient analytical method for different type of edges. As it is known, in general there is no bulk-boundary correspondence for bands
with fragile topology \cite{Song2020}. 
Our method thus provide a direct way to test the appearance of topological edge states within these models.

On the other hand, it is known that the one-electron bands in MATBLG are greatly distorted near the magic angle by the effect of either electron-electron interactions and by the interaction with the substrate \cite{Guinea2020, Zaletel2020, Yacoby2021, Nadj-Perge2021, Wilhelm2020, Dai2021}. To simulate their effect we introduce in the models symmetry breaking terms describing the coupling to the hBN substrate and possible charge density wave (CDW) modulations. This allows us to study the effect of most characteristic non-magnetic perturbations in MATBLG.

Interactions also lead to phases with broken valley and spin symmetries which manifest in the so-called cascade transitions as a function of the doping level \cite{Zondiner2020, Andrei2021}. In this work we do not aim to make predictions on the system phase diagram. We rather assume that valley and spin degrees of freedom can be independently populated and study the resulting spectral and transport properties 
for a given configuration. 

The rest of the manuscript is organized as follows: in Sec.~\ref{sec2}, we discuss the effect of Wannier obstruction and fragile topology of the nearly flat bands of MATBLG in lattice models. We introduce the models from Refs.~\cite{Koshino2018} and \cite{Vishwanath2019}, their relevant symmetries and topology. We implement different non-magnetic symmetry breaking perturbations that can open topological gaps at the Fermi energy. Furthermore, using Wilson loops we obtain the band's Chern number and study the competition among these perturbations in determining the bands topology. 

In Sec.~\ref{sec3}, we generalize the Green function formalism introduced in Ref.~\cite{Alvarado2020} to 2D systems described by lattice models where we only assume finite range hopping amplitudes. To compute 
the boundary Green functions for these models we use an efficient procedure based on the Faddev-Leverrier algorithm \cite{Householder2013}, to be described in a forthcoming publication \cite{forthcoming_paper}. In Sec.~\ref{sec4}, we study the spectral properties at edges of MATBLG starting with the spectral density at a zig-zag boundary showing edge states at the CNP and at the moir\'e gap between the nearly flat bands and the excited ones. Moreover we show the appearance of chiral edge states at the CNP due to sublattice symmetry breaking perturbation in the model of Ref.~\cite{Vishwanath2019} unlike the gaped spectrum for the perturbed model of Ref.~\cite{Koshino2018}. Then we compute the spectral properties at a domain wall for the sublattice symmetry breaking perturbation inducing the appearance of two topological states with the same chirality at opposite minivalleys. In Sec.~\ref{sec5}, we study spectral and transport properties in a three region device where we can modulate the doping levels in those regions, analyzing the influence of topological states at the boundaries of the central region. 

We finally offer, in Sec.~\ref{sec6}, some conclusions summarizing the main results and give an outlook of possible future work taking advantage of the methods developed here. Technical details like the explicit description of the tight-binding models or the recursive GF method to obtain the transport properties in inhomogeneous devices are included in four appendices.

\section{Lattice models for MATBLG}\label{sec2}

Close to the magic angle a lattice model based on the $\pi$-orbitals at the carbon atoms requires more than 10.000 states for each moir\'e unit cell. The idea of minimal lattice models for MATBLG is to construct a basis of well-localized Wannier orbitals purely consisting of flat band states. 
While this idea is extremely attractive, simple two-bands lattice models fail to describe the non trivial topology of the flat bands, which is linked to the fact that they arise from two unperturbed Dirac cones coming from different graphene layers but the same graphene valley, thus carrying the same helicity. 
The topological character of the bands obstructs the construction of minimal 2-bands models based on Wannier orbitals with the correct topology, although they can reproduce the bands dispersion \cite{Koshino2018}. Adding
higher energy trivial bands is necessary to consider all the emergent symmetries of the problem in the hypothesis of
fragile topology \cite{Vishwanath2019}. In spite of these limitations we found it instructive to study first the case of the simplest two band model of Ref. \cite{Koshino2018} and then to extend our study to the six band model of Ref. \cite{Vishwanath2019} that accounts for fragile topology. We shall refer to these two models as two-band-one-valley (2B1V) and six-band-one-valley (6B1V) models respectively.

\subsection{2B1V model}\label{sec2a}
Due to the fact that the flat bands are isolated from the rest of high energy bands, this 2-band minimal model pursues to capture the electronic properties of these bands in a basis of well-localized Wannier orbitals.

\begin{figure}[t]
\includegraphics[width=\columnwidth]{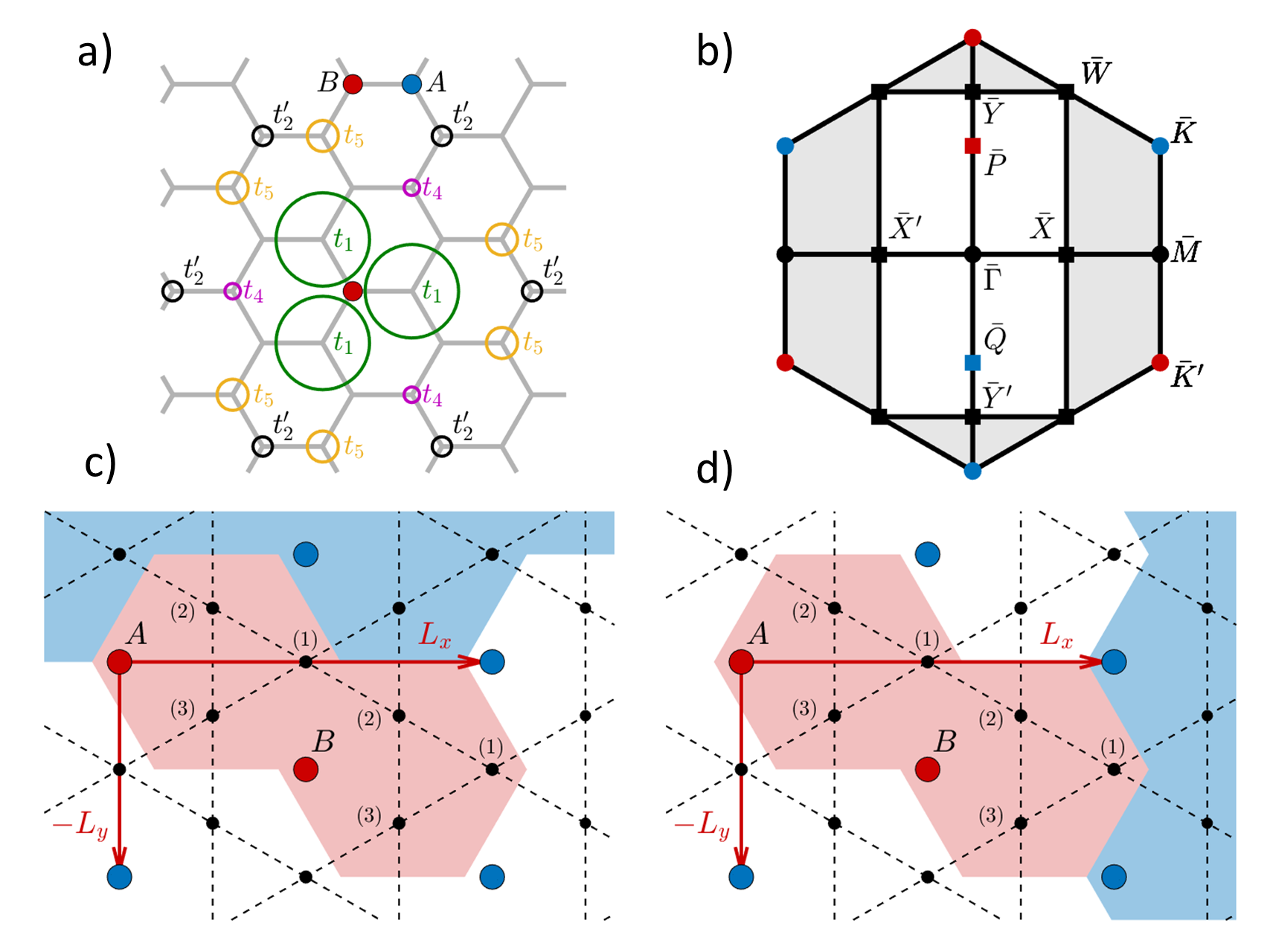}
\caption{Real space representation and corresponding BZs for the 2B1V and 6B1V models.
Panel a) shows the honeycomb lattice and the non-monotonic variation with distance of the hopping integrals ($t_1$, $t_2$, $t_4$, $t_5$) from a lattice site $B$ for the 2B1V model. b) Triangular (grey shaded) and rectangular (white) moir\'e BZ showing the high symmetry points in each one. The overline in the high symmetry points indicates that they belong to the moir\'e BZ. 
Panels c) and d) correspond to the 6B1V model indicating the armchair (panel c) and zigzag (panel d) boundaries in real space.  The $p$-orbitals are centered at the unit cell site red dots $(A,B)$ forming a triangular lattice and the black dots corresponds to the $s$-orbitals placed in a kagome lattice with three sites per unit cell.
The shaded red region in c) and d) corresponds to the doubling of the triangular primitive cell with orthogonal lattice vectors producing the folding of the original triangular BZ into a rectangular one as indicated in panel b).}
\label{fig2}
\end{figure}

The Wannier orbitals in the flat bands have to be centered around the $A = A_1B_2$ and $B = B_1A_2$ sites (see Fig.~\ref{fig2}~a) in the emergent honeycomb moir\'e lattice, where the sub-index refers to different pristine graphene layers in TBG. This description reproduces the observed centering of charge around $AA$ sites in the moir\'e lattice. 

From the localized Wannier description a tight binding Hamiltonian is derived by calculating the hopping integrals between these well localized Wannier orbitals, which behave as $(p_x,p_y)$-orbitals residing each one in one site of the honeycomb lattice. To describe quantitatively the TBG band dispersion it is necessary to take into account hopping elements between an increasing number of neighbours due to rather extended character of the 
Wannier orbitals representing the flat-bands. 
To assure protected Dirac nodes in the flat bands it is necessary to add an additional $\mathbb{Z}_2$ symmetry corresponding to a twofold rotation with a change of valley. 

This model has been intensively used \cite{Recher2020, Koshino2020, Thomale2020, Chichinadze2020} due to its simplicity but, although it exhibits naturally particle/hole symmetry, due to the Wannier obstruction it lacks the characteristic symmetries of the system and leads to moir\'e minivalleys with vanishing net chirality. 

As it was pointed out by Ref.~\cite{Vishwanath2019} this model does not include essential symmetries of the continuum model like $C_2\mathcal{T}$, where $\mathcal{T}$ accounts for time reversal and $C_2$ is a twofold rotation about the z-axis, and consequently, as mention before, has to introduce extra symmetries (i.e. it needs to fine tune certain hoppings to zero) to assure the presence of Dirac nodes. Another example of this failure in capturing the essential symmetries of the continuum model is that a perpendicular electric field $E_z$ opens a gap in this model \cite{Vishwanath2019}.

\begin{table}[t]
  \centering
  \begin{tabular}{|c|l|c|}
    \hline
    \hline
    Parameter & Meaning & Ratio to $t_1$ \\
    \hline
    \hline
    $t_1$ & $(A-B)$ $1^{st}$-nn hopping &  1 \\ 
    $t_2^\prime$ & $(A-A)$ $2^{st}$-nn hopping &  0.208 \\ 
    $t_4$ & $(A-B)$ $2^{nd}$-nn hopping &  0.167 \\ 
    $t_5$ & $(A-B)$ $3^{rd}$-nn hopping &  0.333 \\ 
    \hline
    \hline
  \end{tabular}
  \caption{Hopping amplitudes for the 2B1V model of Ref.\cite{Koshino2018} written in units of the dominant energy scale $t_1 = 1.2$ meV assuming $t_2 \approx it_2^\prime$. }
  \label{table-2B1V}
\end{table}

Following Ref.~\cite{Recher2020} we truncate the hopping terms in the 2B1V model by considering just the four largest ones. As we can see in Table~\ref{table-2B1V}, due to the particular form of the Wannier orbitals the hopping elements do not vary monotonically with distance (i.e. for instance $t_5>t_4$). Notice also that we neglect the $(A-A)\; 1^{st}$-nn hopping $t_3$ which is negligible compared to $t_2$ \cite{Koshino2018}. The relative size of the hopping elements considered is illustrated in Fig.~\ref{fig2}~a).  

Within this approximation the bulk Hamiltonian within the 2B1V model is given by

\bea
&\hat{\mathcal{H}}^{(2)}_\pm(\mathbf{k}) = \bmat 
\pm t_2^\prime g(\mathbf{k}) && h(\mathbf{k}) \\
h(\mathbf{k})^* && \pm t_2^\prime g(\mathbf{k}) \emat,& \nonumber \\ \nonumber \\
&h(\mathbf{k}) =  t_1f_1(\mathbf{k})+t_4f_4(\mathbf{k})+t_5f_5(\mathbf{k}),&
\eea 
where the subindex $\xi=\pm$ indicate the valley, $t_2 \approx it_2^\prime$ and

\bea
g(\mathbf{k}) &=& 2 \sin(\mathbf{k}(2\mathbf{L_2}-\mathbf{L_1})) + 2\sin(\mathbf{k}(2\mathbf{L_1}-\mathbf{L_2}) ) \nonumber \\ &&-2\sin(\mathbf{k}(\mathbf{L_1}+\mathbf{L_2})),  \nonumber \\
f_1(\mathbf{k}) &=& 1+e^{i\mathbf{k} \mathbf{L_1}} +e^{i\mathbf{k} \mathbf{L_2}}, \nonumber \\
f_4(\mathbf{k}) &=& e^{i\mathbf{k} (\mathbf{L_1}+\mathbf{L_2})} + 2\cos(\mathbf{k} (\mathbf{L_1}-\mathbf{L_2})), \nonumber \\
f_5(\mathbf{k}) &=& e^{i2\mathbf{k} \mathbf{L_1}} +e^{i2\mathbf{k}\mathbf{L_2}} +e^{-i\mathbf{k} \mathbf{L_1}} +e^{-i\mathbf{k} \mathbf{L_2}} \nonumber \\ && +e^{i\mathbf{k} (2\mathbf{L_1}-\mathbf{L_2})}+e^{i\mathbf{k} (2\mathbf{L_2}-\mathbf{L_1})} ,
\eea
where the lattice vectors are $\mathbf{L_{1,2}}=L_m(\sqrt{3}/2,\pm 1/2)$ and $L_m = a/(2\sin{\theta/2})$ is the moir\'e lattice vector (given in terms of the graphene lattice parameter $a$ and the twist angle $\theta$). The list of the parameters used are given in Table~\ref{table-2B1V}.

Within this model a gap may be opened by breaking the sublattice symmetry which protects the Dirac nodes with a perturbation of the form $\hat{\mathcal{H}}^\prime_\pm(\mathbf{k}) = \hat{\mathcal{H}}(\mathbf{k})_\pm + \delta_S \sigma_z$. This could be associated either to the effect of the substrate or arising from interactions treated perturbatively in a mean field approximation. In this work we shall consider $\delta_S$ as a phenomenological parameter. 

\begin{figure}[t]
\includegraphics[width=\columnwidth]{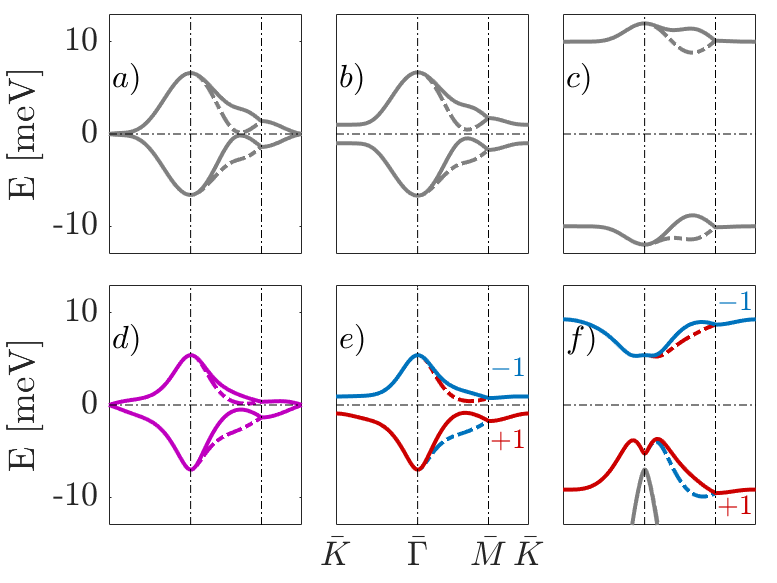}
\caption{Evolution of the nearly-flat bands in the hexagonal BZ for 
the 2B1V (upper pannels) and 6B1V (lower pannels) models due to the sublattice symmetry breaking parameter $\delta_S$, which takes the values 0, 1, 10 meV for each column respectively. Color code: grey bands indicate trivial topological character while colored ones have topological nature. Straight and dashed-dotted lines correspond to different global valleys. Red (blue) is associated to bands with Chern number +1 (-1) on the other hand, magenta means fragile topology bands as in panel d).}
\label{fig3}
\end{figure}

The flat bands in the triangular moir\'e Brillouin zone (BZ) for the 2B1V model are shown in the upper upper panels of Fig.~\ref{fig3}. Full and dashed lines correspond to the two graphene valleys and panels a), b) and c) illustrate the effect of the $\delta_{S}$ parameter opening a gap at the Dirac nodes, which in this model have opposite helicity and thus are trivial in nature.  

\subsection{6B1V model}\label{sec2b}

Ref.~\cite{Vishwanath2019} discusses how to build ``faithful" TB models avoiding the topological obstruction by adding different sets of trivial bands to the flat ones. These models should satisfy the requirements imposed by the relevant symmetries of the system plus the non-trivial topology of the flat bands; and also that the complementary set of trivial bands describe quantitatively the higher energy bands in MATBLG. For that purpose Ref.~\cite{Vishwanath2019} obtains an atomic limit for the sum of the topological flat bands plus the determined set of trivial bands with the constraints of having two isolated bands with the appropriate momentum-space symmetry representations. This results in a basis of $p_z$ and $p_\pm$-orbitals in a triangular lattice $(\tau, p_z)$ and $(\tau, p_\pm)$ respectively, and three $s$-orbitals in a kagome lattice $(\kappa, s)$, using a notation that indicates the lattice (triangular, $\tau$, or kagome, $\kappa$) and the character of the orbitals ($s$ or $p$).

The model includes two $(\tau,p_\pm)-$orbitals sitting on a triangular lattice based on the charge centers in MATBLG at the flat bands placed at the A-A bonds in a triangular moir\'e lattice to reproduce the Dirac nodes of the flat bands, see Fig.~\ref{fig2}~c) and d). To capture the behaviour of the flat bands close to the $\Gamma$ point, however, it is necessary to hybridize these $(\tau, p_\pm)$-orbitals with the rest of the orbitals needed to fulfill the constraints associated to symmetry. 

It should be noticed that recent works question the idea of fragile topology in MATBLG~\cite{Bernevig2020, Song2019}. These works suggest that particle/hole symmetry has to be considered although it is an approximate symmetry inherited from pristine graphene close at charge neutrality. In the models of Ref.~\cite{Vishwanath2019} fragile topology relies exclusively in the $C_2\mathcal{T}$ symmetry and lattice translations which protects the Dirac nodes. According to Refs.\cite{Bernevig2020, Song2019}, with the approximate particle/hole symmetry the system topology becomes stable instead
of fragile. However, no lattice model can reproduce all these symmetries. Still these tight-binding models have been used to predict the correct correlated insulating behaviour \cite{Vafek2019, Yang2020, Nadj_Perge2019, Stauber2018, Chichinadze2020}.


\begin{table}[t]
  \centering
  \begin{tabular}{|c|l|c|}
    \hline
    \hline
    Parameter & Meaning & Ratio to $t_\kappa$ \\
    \hline
    \hline
    
    $t_{p_z}$ & $(\tau, p_z)$ nn hopping &  0.17 \\ 
    $t_{p_\pm}$ & $(\tau, p_\pm)$ nn intra-orbital hopping &  -0.03 \\ 
    $t^+_{p_\pm p_\pm}$ & $(\tau, p_\pm)$ nn inter-orbital hopping & -0.065  \\
    $t^-_{p_\pm p_\pm}$ & $(\tau, p_\pm)$ nn inter-orbital hopping & -0.055  \\
    $t_\kappa$ & $(\kappa, s)$ nn hopping &  1 \\
    $t^\prime_\kappa$ & $(\kappa, s)$ nnn hopping &  0.25 \\
    $t^+_{p_\pm p_z}$ & $(\tau, p_\pm) \times (\tau, p_z)$ nn hopping & 0.095   \\
    $t^-_{p_\pm p_z}$ & $(\tau, p_\pm) \times (\tau, p_z)$ nn hopping & 0.085   \\
    $t^+_{\kappa p_\pm}$ & $(\kappa, s) \times (\tau, p_\pm)$ nn hopping & 0.6   \\
    $t^-_{\kappa p_\pm}$ & $(\kappa, s) \times (\tau, p_\pm)$ nn hopping & 0.2   \\
    
    \hline
    \hline
    
    $\delta_{p_z}$ & $(\tau,p_z)$ chemical potential & -0.2593 \\
    $\delta_{p_\pm}$ & $(\tau,p_\pm)$ chemical potential & -0.3628 \\
    $\delta_{\kappa}$ & $(\kappa,s)$ chemical potential & 0.20 \\
    \hline
    \hline
  \end{tabular}
  \caption{6B1V model parameters written in units of the dominant energy scale $t_\kappa = 27$ meV}
  \label{table-vishwanath}
\end{table}

Although in Ref.~\cite{Vishwanath2019} 5, 6 and 10 bands models have been proposed, in the present work we find it appropriate to concentrate in the 6 bands case. While we are looking for the minimal possible models which accounts for the system fragile topology the reason for this choice is that unlike the 5 bands model, the 6 bands one is formulated in terms of a TB Hamiltonian with usual nearest-neighbours hopping terms instead of quasi-orbitals, which makes it more amenable for applying Green function techniques in real space.  

Details on the 6B1V model as defined on a minimal triangular unit cell can be found in Ref. \cite{Vishwanath2019}. Here we extend it to situations where there could be cell doubling (cd) due to interactions producing charge transfer between neighboring sites \cite{Yacoby2021, Guinea2018, Yazdani2020, Zaletel2020, Chichinadze2020}. 

\begin{figure}[t]
\includegraphics[width=\columnwidth]{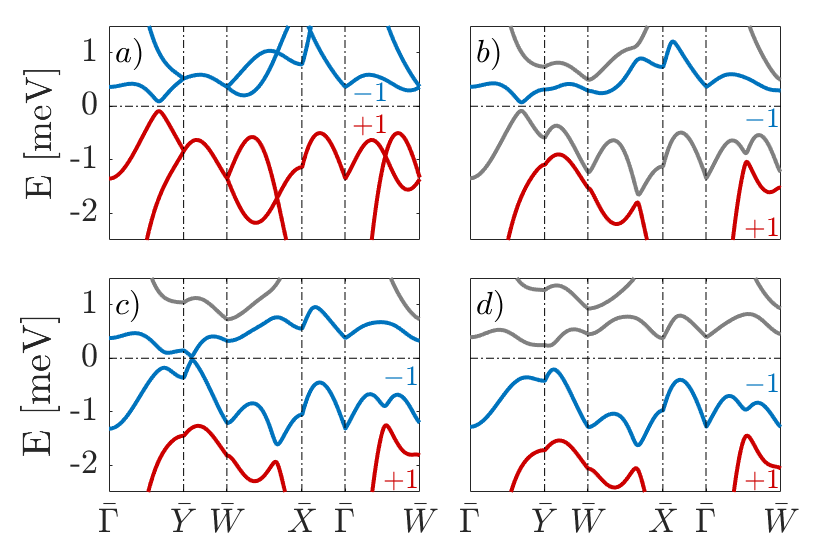}
\caption{Evolution of the nearly-flat bands in the rectangular BZ for one valley in the 
6B1V model due to the translational symmetry breaking potential $\delta_{CDW} = $ 0, 0.3, 0.7, 1 meV for each panel a), b), c), d), respectively, and interaction with the substrate $\delta_S = 0.1$ meV. Color code: grey indicates trivial topological character while colored ones have topological nature. Red (blue) is associated to bands with Chern number +1 (-1).}
\label{fig4}
\end{figure}

This unit-cell doubling leads to an orthogonal lattice with vectors $L_x = \sqrt{3}L_m$ and $L_y = L_m$ (see figure~\ref{fig2}) and the local fermion operators are defined as 
$\hat{\Psi} = (\hat{\Psi}_A \; \hat{\Psi}_B)^T$, with $\hat{\Psi}_{\mu} =
(\hat{\tau}_{p_z, \mu} \; \hat{\tau}_{p_+, \mu} \; \hat{\tau}_{p_-, \mu} \; \hat{\kappa}^{(1)}_{s, \mu}\; \hat{\kappa}^{(2)}_{s, \mu}\; \hat{\kappa}^{(3)}_{s, \mu}\;)^T$,
where $\mu \equiv A,B$ indicates the two sites within the orthogonal cell. Within this basis the 6B1V Hamiltonian adopts the form

\be
\hat{\mathcal{H}}^{(6)}_{cd}(\textbf{k}) = \bmat \hat{H}^{AA} && \hat{H}^{AB} \\
\hat{H}^{AB \dagger} && \hat{H}^{AA} \emat,
\ee
where

\bea
&\hat{H}^{AA} = \bmat H_{p_z}^{AA} +\mu_{p_z} && \hat{C}_{p_\pm p_z}^{AA} && \hat{0} \\
        \hat{C}_{p_\pm p_z}^{AA \dagger} && \hat{H}_{p_\pm}^{AA}+\mu_{p_\pm}\hat{\mathbb{I}}_2 && \hat{C}_{\kappa p_\pm}^{AA} \\
        \hat{0} && \hat{C}_{\kappa p_\pm}^{AA \dagger} && \hat{H}_{\kappa}^{AA} +\mu_\kappa\hat{\mathbb{I}}_3 \emat,& \nonumber \\ \nonumber \\
&\hat{H}^{AB} = \bmat H_{p_z}^{AB} && \hat{C}_{p_\pm p_z, 1}^{AB} && \hat{0} \\
        \hat{C}_{p_\pm p_z, 2}^{AB} && \hat{H}_{p_\pm}^{AB} && \hat{C}_{\kappa p_\pm, 1}^{AB} \\
        \hat{0} && \hat{C}_{\kappa p_\pm, 2}^{AB} && \hat{H}_{\kappa}^{AB} \emat .& 
\eea

The expressions for the matrices $H^{\mu\nu}_{\alpha}$ and $C^{\mu\nu}_{\alpha\beta}$ in terms of the parameters in Table~\ref{table-vishwanath} are given in Appendix~\ref{App_6B1V}. The diagonal elements are set by
$\mu_{p_z} \equiv -6t_{p_z}+\delta_{p_z}$, 
$\mu_{p_\pm} \equiv 3t_{p_\pm}+\delta_{p_\pm}$ and $\mu_{\kappa} \equiv -4(t_\kappa + t_\kappa^\prime)+\delta_{\kappa}$.

In addition, we will consider the following symmetry breaking perturbations

\be
\hat{H}_{cd}^{S} = \delta_S\tau^{p_\pm}_z\sigma_0 \quad,\quad 
\hat{H}_{cd}^{CDW} = \delta_{CDW} \hat{\mathbb{I}}_6 \sigma_z,
\ee
where $\tau^{p_\pm}_z$ is a Pauli matrix acting in the $(\tau,p_\pm)_\mu$-orbitals with $\mu=(A,B)$, $\sigma_0$ is the identity matrix and $\sigma_\mu$ are Pauli matrices which act in the $(A-B)$ doubled unit cell sites space. The $\delta_S$ parameter is equivalent to a sublattice staggered moir\'e potential as pointed out in Ref.~\cite{Carr2019}.

The total Hamiltonian in the 6B1V model is then given by

\be
\hat{\mathcal{H}}^{(6) \prime}_{cd}(\textbf{k}) = \hat{\mathcal{H}}^{(6)}_{cd}(\textbf{k}) + \hat{H}_{cd}^{S} + \hat{H}_{cd}^{CDW}.
\ee

\begin{figure}[t]
\includegraphics[width=\columnwidth]{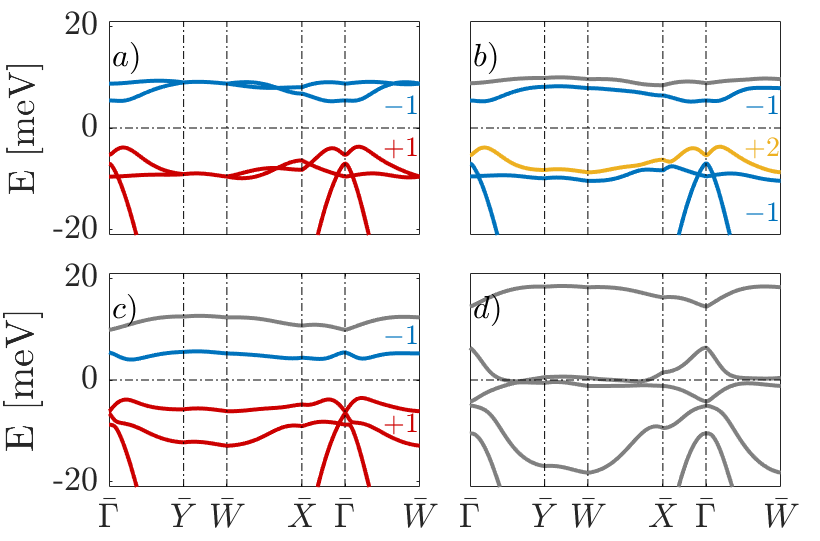}
\caption{Same as in Fig. \ref{fig4} but with the translational symmetry breaking potential $\delta_{CDW} = $ 0, 1, 4, 10.5 meV and strong coupling to the substrate, $\delta_S = 10$ meV. Panel d) shows a gap closing due to band inversion between central Chern bands. Color code: grey bands shows trivial topological character while colored ones have topological nature. Red (blue) is associated to bands with Chern number +1 (-1), yellow in panel b) corresponds to Chern number +2.}
\label{fig5}
\end{figure}

These perturbations can be either associated to the effect of the substrate (i.e. hBN substrate induces sublattice symmetry breaking in the graphene lattice placed on top) or to a CDW arising from interactions (i.e. breaking translational symmetry) allowing us to explore a larger phase space \cite{Wilhelm2020, Dai2021, Guinea2020, Yacoby2021, Yazdani2020, Zaletel2020, Chichinadze2020}.  



On the lower panels of Fig.~\ref{fig3} we show the flat bands in the triangular BZ for the 6B1V model
for increasing values of the gap opening parameter $\delta_S$.
As can be observed, for $\delta_S=0$ the bands within the 2B1V and the 6B1V models are quite similar despite the absent electron/hole symmetry in 6B1V model. For sufficiently large $\delta_S$ their behavior differ substantially. While in the 2B1V model the band minimum continues to appear at $\bar{K}$ and $\bar{K}^\prime$, for the 6B1V model the minimum is transferred into the $\bar{\Gamma}$ point. This behavior (which was also pointed out in Ref.~\cite{Yacoby2021, Guinea2018}) is due to the fact that the states at the $\bar{\Gamma}$ point in the 6B1V model are combination of the $(\tau, p_z)$ and $(\kappa,s)$-orbitals, which are coupled to the perturbed $(\tau, p_\pm)$ ones.

On the other hand, the models differ, as expected, on their topological properties. The different colors in Fig.~\ref{fig3} indicate the bands' Chern number as extracted from the winding of the Berry curvature in the BZ of the corresponding Wilson loops \cite{Alexandradinata2015, Alexandradinata2016, Yang2018, Bradlyn2019}. When $\delta_S \ne 0$ the positive and negative energy flat bands at the CNP in the 6B1V model are characterized by a non-zero Chern number. A change in the global valley induces a change in sign of the Chern number for each band due to time reversal symmetry. The 2B1V bands remain trivial in all cases. 

The effect of the translation symmetry breaking parameter $\delta_{CDW}$ and the combined effect of $\delta_{CDW}$ and $\delta_S$ in the 6B1V model are illustrated in Figs.~\ref{fig4} and \ref{fig5}. In Fig.~\ref{fig4} we show the low energy bands in the orthogonal BZ for weak coupling with the substrate ($\delta_S = 0.1$ meV) and varying $\delta_{CDW}$. The former opens a gap at the CNP and the latter splits each nearly-flat band as shown in panel b), redistributing the topological charge of the original flat bands in a set of splitted trivial and topological bands. The total topological charge at charge neutrality is still $C=+1$. By effect of the competing perturbations a gap closing occurs in panel c) due to a band inversion that produces in panel d) the trivialization of the lower set of flat bands at the CNP.

On the other hand, in Fig.~\ref{fig5} we show the effect of translational symmetry breaking in the case of strong coupling with
the substrate ($\delta_S = 10$ meV). As in Fig.~\ref{fig4} $\delta_{CDW}$ splits the flat bands and redistributes the topological charge between the new set of bands. In this case the lower flat bands are crossed by the set of trivial occupied bands. Although the Chern number at charge neutrality in panel b) is $C=+1$, the redistribution of topological charge is completely different due to the effect of the excited bands which, individually considered, can carry some topological order. For this reason we observe a flat band with $C=+2$. In the subsequent panels we observe different topological transitions: first in panel c) a gap closing between the lower topological bands that set $C=+1$ to the flat band close to the charge neutrality and trivializes the rest of lowest bands and finally in panel d) a band inversion occurs between the flat bands close to the CNP.

The competition between Chern insulating states and CDW in TBLG has been recently analyzed in Refs. \cite{Wilhelm2020, Dai2021,Yang2020}. As we can see in Figs.~\ref{fig4} and \ref{fig5}, starting from a Chern insulating phase the breaking of translational symmetry can lead to a trivial phase, in agreement with these works. 

\section{bGF method for 2D TB models}\label{sec3}

A central aim of the present work is to generalize the bGF technique develop in \cite{Alvarado2020} for 1D models to the case of 2D lattice models, particularly to the ones describing bulk MATBLG by a $N\times N$ Hamiltonian in reciprocal space $\hat{\mathcal{H}}(\textbf{k})$. 

\begin{figure}[t]
\includegraphics[width=\columnwidth]{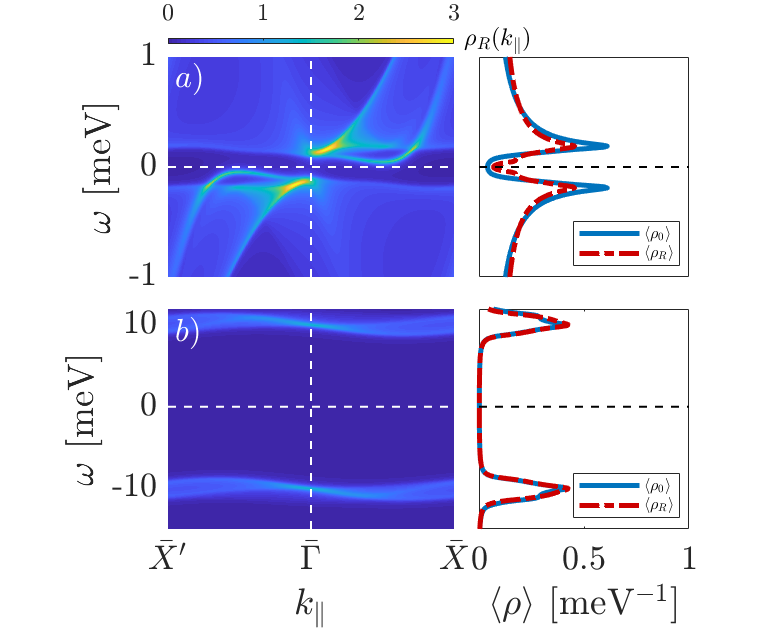}
\caption{Open AC boundary LDOS for the 2B1V model showing trivial edge states under the effect of the breaking of sublattice symmetry $\delta_{S}$. Left column: spectral density for a right boundary. Right column: integrated LDOS where straight (dot-dashed) line represents bulk (right boundary) LDOS. Panel a): Weak coupling limit $\delta_{S} = 0.1$ meV where a gap opens with a minimal distorsion of the bands. It exhibits trivial edge states. Panel b): Strong coupling limit $\delta_{S} = 10$ meV where the bands are greatly distorted and the trivial edge states disappear.}
\label{fig6}
\end{figure}

In order to compute the bGF we decompose the momenta into parallel and perpendicular components $\textbf{k} = (k_\parallel, k_\perp)$ relative to the boundary direction that we consider (in higher dimensional models the parallel momentum component would be itself a vector $\textbf{k}_\parallel$). The bulk Hamiltonian periodicity in both directions is set by $(L_\parallel, L_\perp)$, such that $\hat{\mathcal{H}}(
\textbf{k}+2\pi\textbf{u}_\perp/L_\perp)=\hat{\mathcal{H}}(\textbf{k})$, where $\textbf{u}_\perp$ is the unitary vector in the perpendicular direction. Thus, for a model with a triangular lattice like the 6B1V model 
we have to double the primitive cell to obtain an orthogonal decomposition of the momentum. Using this periodicity, the Hamiltonian can be expanded in a Fourier series, $\hat{\mathcal{H}}(\textbf{k})=\sum_n \hat{\mathcal{V}}_n(k_\parallel) e^{ink_\perp L_\perp}$, where $n$ is the number of neighbours and Hermiticity implies $\hat{\mathcal{V}}_{-n}=\hat{\mathcal{V}}^{\dagger}_n$. 

The advanced bulk GF is then defined as
\begin{equation}
\hat{G}^A(\textbf{k},\omega) = \left[(\omega-i0^+)\hat{\mathbb{I}} - \hat{\mathcal{H}}(\textbf{k})\right]^{-1},
\end{equation}
where the $N\times N$ matrix structure is indicated by the hat notation.
Fourier transforming along the perpendicular direction, the GF components are given by
\be\label{GR11}
\hat{G}^A_{jj'}(k_\parallel, \omega) = \frac{L_\perp}{2\pi} \int\limits_{-\pi/L_\perp}^{\pi/L_\perp}dk_\perp  e^{i(j-j')k_\perp L_\perp}\, \hat{G}^A(k_\parallel,k_\perp, \omega) ,
\ee
where $j$ and $j'$ are lattice site indices.
By the identification $z = e^{ik_\perp L_\perp}$, this integral is converted into a complex contour integral, 

\be\label{G1}
\hat{G}^A_{jj'}(k_\parallel,\omega) = \frac{1}{2\pi i} \oint\limits_{|z|=1} \frac{dz}{z} z^{j-j'} \hat{G}^A(k_\parallel,z,\omega).
\ee

\begin{figure}[t]
\includegraphics[width=\columnwidth]{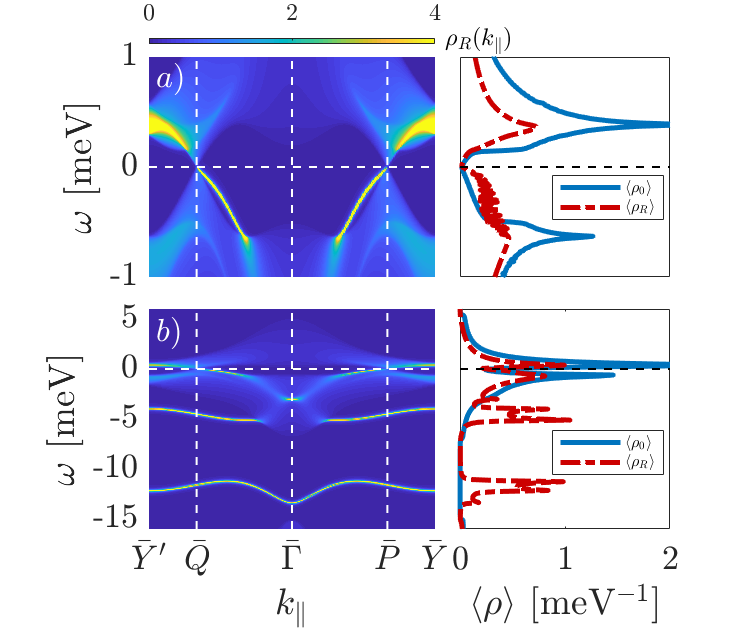}
\caption{Open ZZ boundary LDOS for the 6B1V model without symmetry breaking perturbations. 
 As in Fig. \ref{fig6} the left column corresponds to the spectral density for a right boundary while the right column shows the integrated LDOS where straight (dot-dashed) line represents bulk LDOS $\langle \rho_0 \rangle$ (right boundary LDOS $\langle \rho_R \rangle$). The upper and lower panels show the same results but in different energy scale. In panel a) we can observe trivial edge states inherit from pristine monolayer graphene appearing within the nearly-flat bands. Panel b) illustrates the appearance of moir\'e edge states in the gap between nearly-flat and the lower energy bands.}
\label{fig7}
\end{figure}
\noindent

Further simplification can be obtained by introducing the roots $z_n(k_\parallel,\omega)$ of the secular polynomial in the complex-$z$ plane,

\bea\label{polysec}
P(k_\parallel,z,\omega) &=& \mbox{det}\left[\omega \hat{\mathbb{I}} - \hat{\mathcal{H}}(k_\parallel,z)\right] \nonumber \\
&=& \frac{c_m}{z^m}\prod_{n=1}^{2m} \left[z-z_n(k_\parallel,\omega)\right],
\eea
where $m$ is the highest order of the polynomial and $c_m$ is the highest order coefficient. In terms of these roots the contour integral in Eq.~\eqref{G1} can be written as a sum over the residues of all roots inside the unit circle

\begin{equation}
\hat{G}^A_{jj'}(k_\parallel,\omega)  = \sideset{}{'}\sum_{|z_n|<1} \frac{z_n^{p}\hat{A}(k_\parallel,z_n,\omega) }{c_m\prod_{l\ne n}
\left(z_n - z_l\right)},
\label{residues}
\end{equation}
where $p = j-j'+m-m^{\prime}-1$ and $z^{-m^\prime}\hat{A}(k_\parallel,z,\omega)$ is the adjugate matrix of $\left[\omega \hat{\mathbb{I}} - \hat{\mathcal{H}}(k_\parallel,z)\right]$ where all the poles at zero were taken out of $\hat{A}$ as a common factor in $z^{-m^\prime}$. Finally, $\sum^{'}$ means that if $p=-1$, then we include $z_n=0$ as a pole to take into account in the sum of residues (i.e. in the non local GF components). When $p<-1$ higher order poles at zero appear in the sum of residues. To simplify we can take advantage of the residue theorem to avoid these poles and compute the integral as
\begin{equation}\label{adjugate_absplus}
\hat{G}^A_{jj'}(k_\parallel,\omega)  = -\sum_{|z_n|>1} \frac{z_n^{p}\hat{A}(k_\parallel,z_n,\omega) }{c_m\prod_{l\ne n}
\left(z_n - z_l\right)}.
\end{equation}

\begin{figure}[t]
\includegraphics[width=\columnwidth]{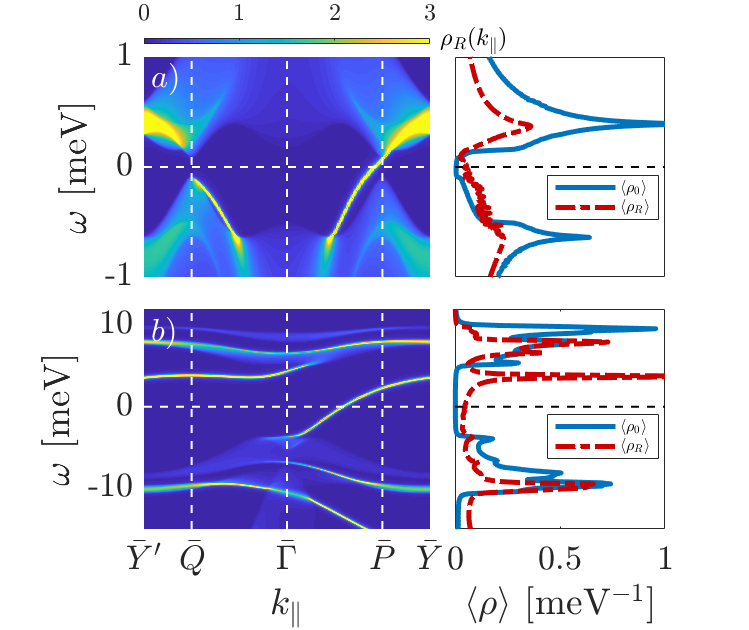}
\caption{Same as Fig. \ref{fig7} but under the effect of the coupling with the substrate $\delta_S$ and broken translation symmetry $\delta_{CDW}$. Panel a): Weak coupling with the substrate ($\delta_S = 0.1$ meV) where interactions open a gap without distorting the bands. It shows the appearance of topological edge state between the flat bands at the $\bar{P}$ minivalley. Panel b): Strong coupling with the substrate ($\delta_S = 10$ meV) where interactions distort the bands and the breaking of translational symmetry with $\delta_{CDW} = 1$ meV splits both flat bands. One can observe topological edge states between the flat bands and another topological state with opposite chirality between the splitted flat band and the lower energy bands due to breaking of translation symmetry.}
\label{fig8}
\end{figure}

In \cite{forthcoming_paper} we describe an efficient method, based on Fadeev-LeVerrier algorithm to construct the secular polynomial and the adjugate matrix which are needed to evaluate the GF components in Eq.~\eqref{residues}. On the other hand, in Appendix~\ref{App_bGF} we give details on the method to obtain the bGF for the system with an open boundary and to get the transport properties for junctions between different TBLG regions.


\section{Spectral properties at edges}\label{sec4}

The bGF $\hat{\mathcal{G}}_{L,R}$ of the semi-infinite system obtained as described in Appendix~\ref{App_bGF} allows us to analyze the spectral properties at the edges of MATBLG, encoded in the spectral densities $\rho_{L,R}(k_\parallel,\omega) = \mbox{Im} \hat{\mathcal{G}}_{L,R}(k_\parallel,\omega)/\pi$ and the LDOS $\bra \rho_{L,R}(\omega) \ket = \int dk_\parallel/\Omega_{k_\parallel} \rho_{L,R}(k_\parallel,\omega)$, where $\Omega_{k_\parallel} = 2\pi/L_\parallel$ accounts for the limits of integration. 

In Fig.~\ref{fig6} we illustrate the spectral properties at an armchair edge for the 2B1V model in the 
weak ($\delta_{S}=0.1$ meV) (a) and strong ($\delta_{S}=10$ meV) (b) coupling limits. In the AC boundary the $\bar{P}$ and $\bar{Q}$ minivalleys are projected onto the $\bar{\Gamma}$ point.
As expected, no chiral states appear when opening a gap in the 2B1V model. Only trivial states can be observed in case (a). In the strong coupling limit the edge states just merge with the continuum.

\begin{figure}[t]
\includegraphics[width=\columnwidth]{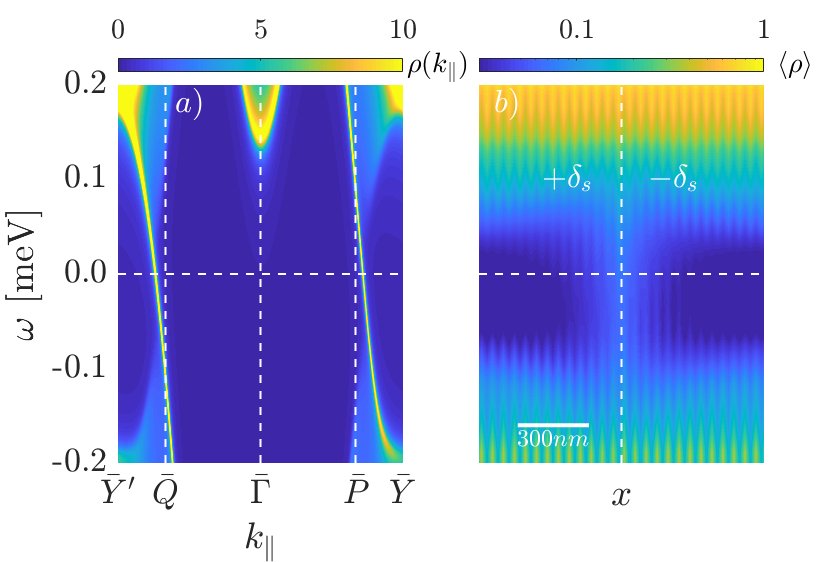}
\caption{Topological domain wall with ZZ orientation in the 6B1V model. This is produced
by a change of sign of $|\delta_S| = 0.1$ meV across the junction and we also include translational symmetry breaking with $\delta_{CDW} = 0.5$ meV. Panel a) shows the local spectral density at the domain wall where a chiral state for each mini-valley with the same chirality can be observed. Panel b) shows the integrated LDOS in log-scale along the domain wall where the chiral states are localized around the center with a characteristic penetration length of $\approx 200nm$. The vertical white dashed line represents the domain wall frontier where the sign of $\delta_S$ changes. The density of states are measured in [meV$^{-1}$] units.
}
\label{fig9}
\end{figure}

The edge spectral properties are radically different in the 6B1V model.
Due to the contribution of the set of higher energy bands in the 6B1V model we can distinguish two types of edge states in Fig.~\ref{fig7}: the ones inherited from the pristine graphene appearing at the charge neutrality point (CPN) at each minivalley ($\bar{Q}$ and $\bar{P}$) in panel (a) and the ones that emerge entirely from the moir\'e structure at the gap between the flat bands and the excited ones \cite{Koshino2020} (see panel b). One of these moir\'e edge states merges with the flat bands at $\bar{\Gamma}$ and the other one is gaped and nearly dispersionless. This gaped edge state also appears at the AC boundary and might be related to higher order topology \cite{Park2019, Park2021, Liu2021}.

A gap opening at the CNP allows us to confirm the topological character of the nearly flat bands. In Fig.~\ref{fig8}~a) for a right ZZ boundary a trivial edge state persists around the $\bar{Q}$ minivalley and there appears a chiral edge state around $\bar{P}$ minivalley in the weak coupling limit with $\delta_S = 0.1$ meV. In addition, the left boundary shows a topological edge state with opposite chirality at $\bar{Q}$. 

\begin{figure}[t]
\includegraphics[width=\columnwidth]{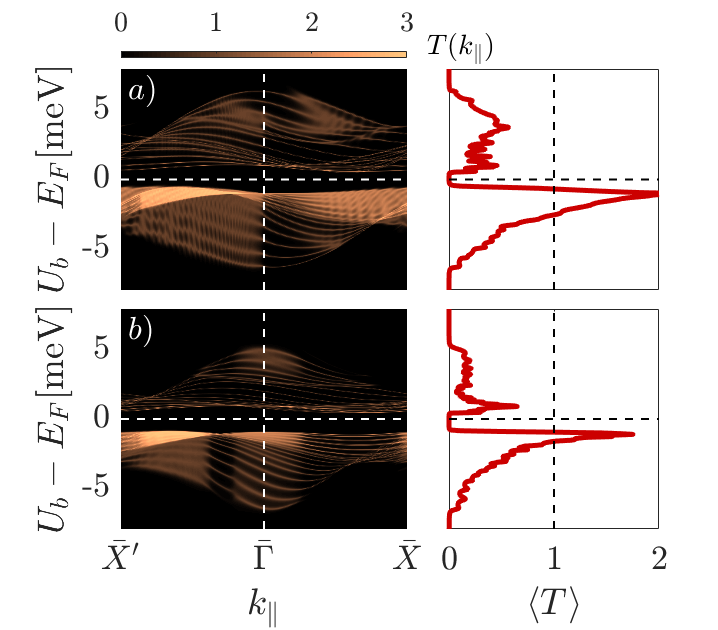}
\caption{Two-terminal transport for a three region junction device with AC orientation as a function of the height of the potential barrier $U_b$ in the central region for a) 2B1V and b) 6B1V models. In both models we take $\delta_S = 1$ meV. Left column: momentum resolved transmission of the junction. Right column: Solid red line represents the average transmission. The length of the central region is $\lambda_b \approx 180$ nm and the Fermi energy is kept at $E_F = -1$ meV on the lateral regions.}
\label{fig10}
\end{figure}

In panel b) we can analyze the effect of $\delta_{CDW}$ in the strong coupling limit,  $\delta_S = 10$ meV. In the first place, there appears a chiral edge state at charge neutrality between the new band minima around $\bar{\Gamma}$ and thus it loses the minivalley character which was present in panel a). As we pointed out in Fig.~\ref{fig5}~b), at CNP the occupied bands have $C=+1$. On the other side, the gap between the lower flat band and the rest of lower energy bands has $C=-1$. For that reason we can observe another chiral edge state with opposite chirality with respect to the one at charge neutrality. Notice that changing the global valley degree of freedom we obtain a similar picture for each boundary where the topological edge states have the opposite chirality and appears at the opposite minivalley (results not shown)

\begin{figure*}[t]
\includegraphics[width=1\textwidth]{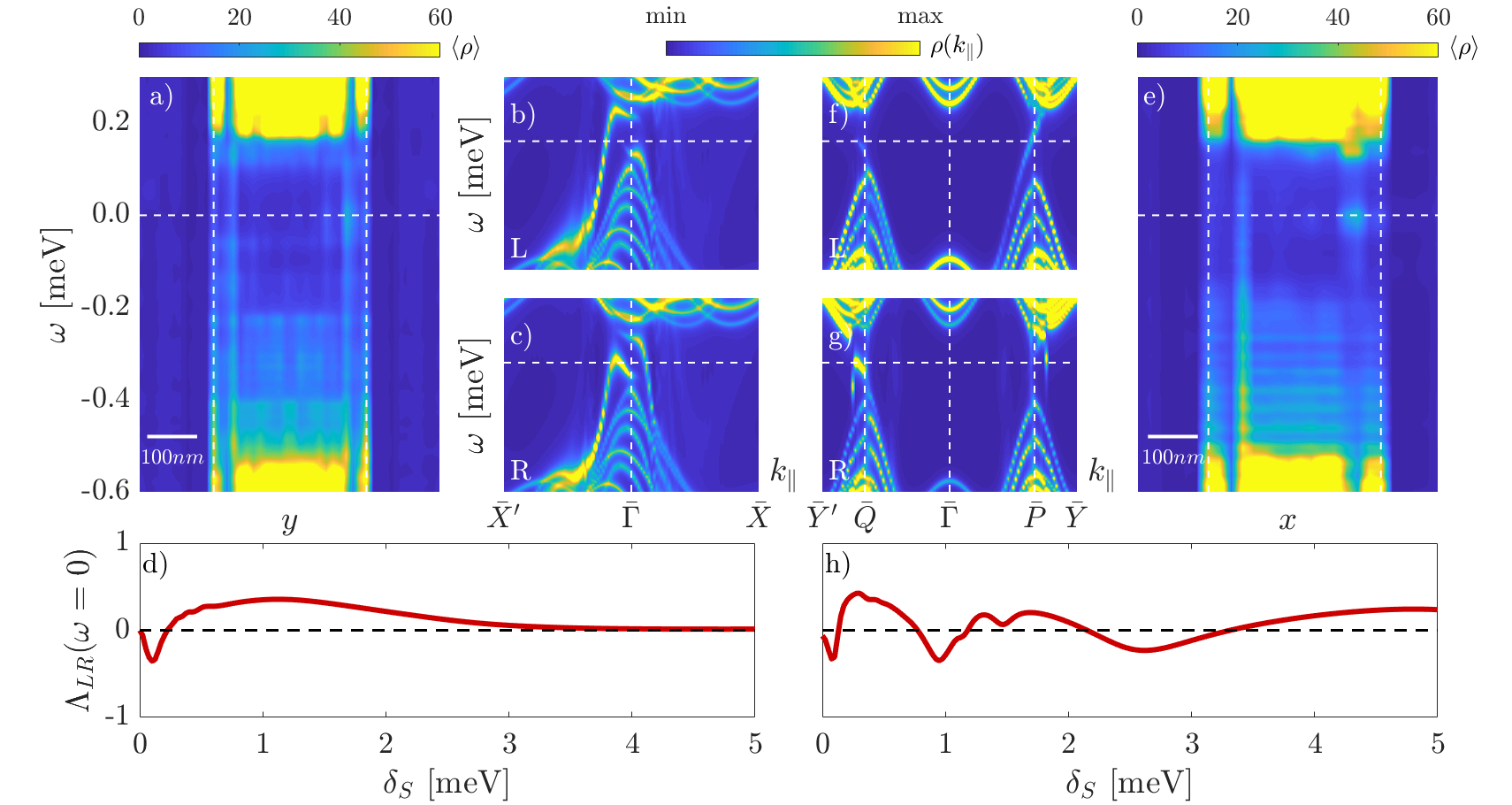}
\caption{Finite size effects in the local spectroscopy for a three region junction N-CNP-N with $\delta_S= 0.1$ meV for both AC (left panels) and ZZ (right panels) boundaries. Panel a) and e): LDOS across the central region of the junction for AC and ZZ boundaries respectively. We can observe anticrossings at low energies due to the hybridization of the edge states located at the ends of the central region. The central panels show the spectral density at both ends of the central region of each junction, upper (lower) ones shows the left (right) end in arbitrary units. Panels d) and h) show the LDOS asymmetry dimensionless factor $\Lambda_{LR}$ between the region ends at the CNP as a function of the sublattice symmetry breaking potential $\delta_S$. The length of the central region is $\lambda_b \approx 270$ nm and its doping level is kept at the CNP, the lateral N regions are heavily doped with $E_F = -140$ meV.}
\label{fig11}
\end{figure*}

In Fig.~\ref{fig9} we show the spectral properties at a domain wall formed by an abrupt change in the sign of the coupling with the substrate $\delta_S$ as illustrated in Fig.~\ref{fig1}~b). As can be observed in panel a) there appear two chiral edge states at different minivalleys propagating in the same direction. These correspond to  states on the two sides of the domain wall which do not annihilate as they ``live" at different minivalleys. Notice that the chirality of the valley polarized edge states depends on the sign of the sublattice perturbation $\delta_S$ and on the general valley.

In panel b) we show the LDOS as a function of the distance from the domain wall located at the center of the plot. It illustrates how the chiral edge state inside the gap decays on a scale of 100 nm and the bulk gap is reconstructed.

\section{Transport at N-N'-N junctions}\label{sec5}

In this section we analyze two terminal transport in different types of MATBLG junctions.
In Appendix.~\ref{App_rGF} we extend the GF method to take into account a non-uniform potential profile in the central region, as in the situation depicted in Fig.~\ref{fig1}~a).

The two-terminal conductance properties which are predicted by the 2B1V and 6B1V models in the situation of Fig.~\ref{fig1}~a) are illustrated in Fig.~\ref{fig10} as a function of the height of the barrier, $U_b$, which defines the doping level in the central region. The lateral regions are kept at $E_F = -1$ meV. For both models we include a finite gap opening parameter of the same size ($\delta_S$ = 1 meV). As can be observed, in Fig.~\ref{fig10} there is a qualitative and quantitative agreement between the two models, which reflects the fact that the topological character of the bands in the 6B1V model does not play a role in this case. 
In agreement with results of Ref.~\cite{Recher2020} for an abrupt barrier the momentum resolved transmission exhibits oscillations due to finite size effects. 
This spectral density oscillations are linked to the finite length of the central region $\lambda_B$.
We can also observe that the normalized total transmission exhibits a maximum at the Van Hove singularity. A lower transmission is obtained when the barrier induces injection of states through the valence band.


The topological properties, however, do play a role in the case of N-N'-N junctions where N' is kept at the CNP and the side regions are heavily doped. The outer regions in this junction exhibit smaller DOS than the flat bands, differing from the metallic behaviour of heavily doped pristine graphene. The spectral properties for this type of junctions are illustrated in Fig.~\ref{fig11}. We consider two different orientations of the junctions with respect to the moir\'e lattice: AC (panels a-d) and ZZ (panels e-h).
As illustrated by these results, the orientation of the junctions leads to substantial differences even when the gate potential profile is kept constant. 

In the case of the armchair orientation we observe unexpected but mild left-right (LR) asymmetry in the LDOS. This is due to the sublattice symmetry breaking by $\delta_S$ which acts in the directional $(\tau,p_\pm)$-orbitals. Due to the large penetration depth of the chiral edge states located at the ends of the central region we observe minigaps associated to hybridization of these states. The spectral densities of the boundaries show the hybridized opposite chirality end states. To analyze the evolution of the LR asymmetry as a function of $\delta_S$ we define

\be
\Lambda_{LR}(\omega) = \frac{\bra \rho_L(\omega) \ket-\bra \rho_R(\omega) \ket}{\bra \rho_L(\omega) \ket+\bra \rho_R(\omega) \ket},
\ee
which corresponds to a LDOS LR asymmetry dimensionless factor. Panel d) shows that in the limit $\delta_S \rightarrow 0$ the AC junction does not show any LR asymmetry because without perturbations both AC boundaries are equivalent. Valley symmetry does not lift this LDOS asymmetry and only induces an inversion of the spectral density
in $k_{\parallel}$ (i.e. changing the chirality and minivalley of the topological edge states). The step-like increase of the LDOS at $\omega \approx -0.15$ meV is due to the top of valence band. The bottom of the conduction band, placed at $\omega \approx 0.2$ meV.

On the other hand, ZZ junctions exhibit a much stronger LR asymmetry which survives even in the limit $\delta_S \rightarrow 0$, see panel h). Within the 6B1V model this is due to the definition of the unit cell and the effect of kagome orbitals at the boundaries, as can be seen in Fig.~\ref{fig2}~d). Left (right) boundaries are defined on triangular (kagome) sites and for that reason both boundaries are naturally non-equivalent. This unperturbed limit for the LR asymmetry relies in the fact the the charge centers are displaced from the triangular lattice sites in the direction perpendicular to the ZZ edge by effect of the kagome sites. Nevertheless, this is not exclusively a feature of the 6B1V model but a general property of MATBG with Wannier orbitals resembling a fidget spinner \cite{Koshino2018, Guinea2018, po2018}.

In addition, as for ZZ edges minivalleys are not coupled and opposite chirality states are placed at different minivalleys, the hybridization between chiral states is also asymmetric inducing anticrossings at different places in the BZ. Panel f) shows an interplay between the chiral states at $\bar{P}$ and the trivial ZZ edge state around $\bar{Q}$ at the left boundary, which hybridizes with the chiral states at the right boundary at $\bar{Q}$, see panel g). 

\begin{figure}[t]
\includegraphics[width=\columnwidth]{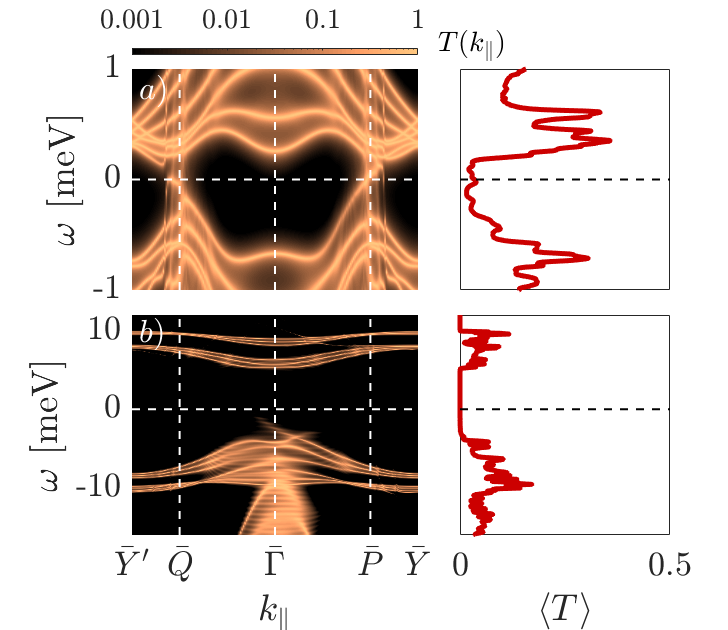}
\caption{Two-terminal transport properties for a three region junction N-CNP-N with ZZ orientation. Panel a): Weak coupling limit with the substrate $\delta_S= 0.1$ meV. Panel b): Strong coupling limit with the substrate $\delta_S= 10$ meV plus breaking of translational symmetry with $\delta_{CDW}= 1$ meV. Left column: spectral representation of the transmission at the junction in log-scale. Right column: solid red line represents the average transmission of the nearly-flat bands. The length of the central region is $\lambda_b \approx 60$ nm and its Fermi energy is kept at the CNP, while the normal regions N are heavily doped with $E_F = -140$ meV.}
\label{fig12}
\end{figure}

Finally in Fig.~\ref{fig12}~a) we illustrate the transport properties for a gate voltage potential profile as in Fig.~\ref{fig11} but with a reduced length of the central region to $\lambda_b \approx 60$ nm in order to obtain a larger overlap between the chiral edge states at the ends of this narrow central region. We observe in this case a small but finite transmission in the gap region which can be associated with the chiral states hybridization. On the other hand, for the strong coupling case the overlap diminishes because the penetration length decays with $\delta_S$ and thus the transmission in the gap region becomes negligible, see Fig.~\ref{fig12}~b).


\section{Conclusions and outlook}\label{sec6}

In this work we have analyzed electronic and transport properties of MATBLG junctions using lattice models. We have considered both the 2B1V model of Ref. \cite{Koshino2018} and the 6B1V model of Ref. \cite{Vishwanath2019}, the last one accounting for fragile topology. For that purpose we have implemented a boundary Green function approach by extending the method developed in Ref. \cite{Alvarado2020} to 2D lattices with hopping elements between arbitrary distant neighbors. 

Using this approach we have analyzed the spectral properties at different types of edges showing the appearance of topological states for the 6B1V model in the presence of symmetry breaking perturbations. We have also analyzed the LDOS at a domain wall defined by the sublattice symmetry breaking potential showing the coexistence of topological edge states at different minivalleys with the same chirality. This analysis allows us to characterize the penetration length of the chiral states.

Regarding the transport properties between MATLBG regions doped at the flat bands through a barrier, we have found that these are not sensitive to topology and thus both the 2B1V and the 6B1V models give qualitatively similar results. In contrast, for a three region junction where the central one is fixed at the CNP and the lateral ones are heavily doped, transport and LDOS are determined by the hybridization of chiral states running in opposite directions at the interfaces. As we have shown, an asymmetry in 
the LDOS between the left and right interface appears which could allow one to distinguish the orientation of the junction with respect to the moir\'e lattice.   

As an outlook, we foresee several possible applications of the method developed in the present work. On the one hand, an extension to a Bogoliubov de Gennes formulation would allow to study Josephson and Andreev transport in MATBLG junctions including superconducting correlations \cite{Pablo2008, Pablo2009, Pablo2010}, in line with recent experiments \cite{Rodan-Legrain2020}. 
This type of calculations could help to identify the symmetry of the order parameter in superconducting MATBLG which is at present an open issue attracting great interest \cite{Thomale2020, Scalettar2018, Ray2019, Christos2020, Peri2021}. Similar calculations have been implemented in the past by some of the authors for the case of pristine graphene layers in proximity with unconventional and topological superconductors \cite{Casas2019}. Another extension of our work could be oriented to analyze the possible higher-order topological insulating (HOTI) states in MATBLG and the appearance of corner states associated to the gapped moir\'e bound states \cite{Yang2019, Wieder2018, Park2019, Liu2021, Park2021}.
 
\begin{acknowledgements}
We acknowledge fruitful discussions with L. Brey, P. Burset, Ch. de Beule, P. Recher and T. Stauber and thank all them for useful comments on this manuscript. This project has been funded by the Spanish MICINN through Grant No.~FIS2017-84860-R; and by the Mar{\'i}a de Maeztu Programme for Units of Excellence in n Research and Development Grant No.~MDM-2014-0377.
\end{acknowledgements}


\appendix

\begin{widetext}
\section{6B1V Hamiltonian with unit cell doubling}\label{App_6B1V}
As defined in the main text, for the 6B1V model we consider a doubled unit cell leading to an orthogonal lattice with vectors $L_x = \sqrt{3} 
L_m$ and $L_y = L_m$ and local fermion operators 
$\hat{\Psi} = (\hat{\Psi}_A \; \hat{\Psi}_B)^T$, with $\hat{\Psi}_{\mu} =
(\hat{\tau}_{p_z, \mu} \; \hat{\tau}_{p_+, \mu} \; \hat{\tau}_{p_-, \mu} \; \hat{\kappa}^{(1)}_{s, \mu}\; \hat{\kappa}^{(2)}_{s, \mu}\; \hat{\kappa}^{(3)}_{s, \mu})^T$,
where $\mu \equiv A,B$ indicates the two sites within the doubled unit-cell. The 6B1V Hamiltonian in this basis adopts the general form 

\bea 
&\hat{\mathcal{H}}^{(6)}_{cd}(\textbf{k}) = \bmat \hat{H}^{AA} && \hat{H}^{AB} \\
\hat{H}^{AB \dagger} && \hat{H}^{AA} \emat ,&
\eea
where


\be
\hat{H}^{AA} = \bmat H_{p_z}^{AA} +\mu_{p_z} && \hat{C}_{p_\pm p_z}^{AA} && \hat{0} \\
        \hat{C}_{p_\pm p_z}^{AA \dagger} && \hat{H}_{p_\pm}^{AA}+\mu_{p_\pm}\hat{\mathbb{I}}_2 && \hat{C}_{\kappa p_\pm}^{AA} \\
        \hat{0} && \hat{C}_{\kappa p_\pm}^{AA \dagger} && \hat{H}_{\kappa}^{AA} +\mu_\kappa\hat{\mathbb{I}}_3 \emat \quad , \quad 
\hat{H}^{AB} = \bmat H_{p_z}^{AB} && 
        \hat{C}_{p_\pm p_z, 1}^{AB} && \hat{0} \\
        \hat{C}_{p_\pm p_z, 2}^{AB} && \hat{H}_{p_\pm}^{AB} && \hat{C}_{\kappa p_\pm, 1}^{AB} \\
        \hat{0} && \hat{C}_{\kappa p_\pm, 2}^{AB} && \hat{H}_{\kappa}^{AB} \emat .
\ee

We now define all the contributions to the total Hamiltonian classified for its orbital character. First the nearest-neighbor standard hopping for $(\tau, p_z)$-orbitals in a cd basis


\be
H_{p_z}^{AA} = t_{p_z}(\phi_{0\bar{1}} + \phi_{01}) \quad , \quad H_{p_z}^{AB} = t_{p_z}(1 + \phi_{0\bar{1}}(1 + \phi_{10}) + \phi_{10}),
\ee
where $\phi_{10} = e^{-i k_x L_x}$ and $\phi_{01} = e^{-i k_y L_y}$ with $\phi_{\bar{ij}} = \phi_{ij}^\dag$. 

For the triangular lattice $(\tau,p_\pm)-$orbitals  we have the following intra-orbital terms


\be
\hat{H}_{p_\pm, 0}^{AA} = t_{p_\pm}(\phi_{0\bar{1}} + \phi_{01})\hat{\mathbb{I}}_2 \quad , \quad
\hat{H}_{p_\pm, 0}^{AB} = t_{p_\pm}(1 + \phi_{0\bar{1}}(1 + \phi_{10}) + \phi_{10})\hat{\mathbb{I}}_2,
\ee
and the inter-orbital terms

\bea 
&\hat{C}_{p_\pm p_\pm}^{AA} = \bmat 0 && t^+_{p_\pm p_\pm}\phi_{0\bar{1}} + t^-_{p_\pm p_\pm}\phi_{01} \\ 
t^+_{p_\pm p_\pm}\phi_{01} + t^-_{p_\pm p_\pm}\phi_{0\bar{1}} && 0 \emat,& \nonumber \\
&\hat{C}_{p_\pm p_\pm}^{AB} = \bmat 0 &&  t^+_{p_\pm p_\pm}(w+w^\prime \phi_{10}) + t^-_{p_\pm p_\pm}(w^\prime\phi_{0\bar{1}} +w\phi_{0\bar{1}}\phi_{10})\\  
t^+_{p_\pm p_\pm}(w^\prime\phi_{0\bar{1}}\phi_{10} +w\phi_{0\bar{1}}) +
t^-_{p_\pm p_\pm}(w^\prime+w \phi_{10}) && 0 \emat&, 
\eea  
where $w=e^{i 2\pi/3}$ and $w^\prime$ is the complex conjugate of $w$. This phase accounts the different orientation of $(\tau,p_\pm)$-orbitals. We can change valley with the transformation $w \rightarrow w^\prime$. The final expression of the total Hamiltonian for this subspace has the form

\be
\hat{H}_{p_\pm}^{AA} = \hat{H}_{p_\pm, 0}^{AA} + \hat{C}_{p_\pm p_\pm}^{AA} \quad,\quad
\hat{H}_{p_\pm}^{AB} = \hat{H}_{p_\pm, 0}^{AB} + \hat{C}_{p_\pm p_\pm}^{AB},
\ee
which contains the fundamental symmetries that protect the Dirac nodes in the triangular lattice. Next we consider the hopping terms concerning the kagome lattice $(\kappa,s)$-orbitals which include first and second nearest neighbour contributions within this subspace in the form

\bea 
&\hat{H}_\kappa^{AA} = t_\kappa \bmat 0 && 1 && 1 \\ 1 && 0 && 1+\phi_{0\bar{1}}\\ 1 && 1+\phi_{01} && 0 \emat+ t^\prime_\kappa
        \bmat 0 && \phi_{01} && \phi_{0\bar{1}} \\ \phi_{0\bar{1}} && 0 && 0 \\
        \phi_{01} && 0 && 0\emat,& \nonumber \\ \nonumber \\
&\hat{H}_\kappa^{AB} = t_\kappa \bmat 0 && 1 && \phi_{0\bar{1}} \\
        \phi_{0\bar{1}} \phi_{10} && 0 && 0 \\
        \phi_{10} && 0 && 0 \emat + t^\prime_\kappa
        \bmat 0 && \phi_{0\bar{1}} && 1 \\ \phi_{10} && 0 && \phi_{0\bar{1}}(1+\phi_{10}) \\
        \phi_{10}\phi_{0\bar{1}} && 1+\phi_{10} && 0\emat.&
\eea


On the other hand, the coupling terms between the $(\tau,p_z)$ and the $(\tau,p_\pm)$ -orbitals are given by

\bea 
&\hat{C}_{p_\pm p_z}^{AA} = i\bmat -t^+_{p_\pm p_z}\phi_{0\bar{1}} + t^-_{p_\pm p_z}\phi_{01} \\ t^+_{p_\pm p_z}\phi_{01} - t^-_{p_\pm p_z}\phi_{0\bar{1}} \emat^T \quad,\quad 
\hat{C}_{p_\pm p_z, 1}^{AB}  = i\bmat -t^+_{p_\pm p_z}(w+w^\prime \phi_{10}) + t^-_{p_\pm p_z}(w^\prime\phi_{0\bar{1}} + w\phi_{0\bar{1}}\phi_{10}) \\ t^+_{p_\pm p_z}(w\phi_{0\bar{1}} + w^\prime\phi_{0\bar{1}}\phi_{10})-t^-_{p_\pm p_z}(w^\prime +w \phi_{10}) \emat^T,& \nonumber \\ \nonumber \\
&\hat{C}_{p_\pm p_z, 2}^{AB} = i\bmat t^+_{p_\pm p_z}(w\phi_{0\bar{1}}+w^\prime \phi_{0\bar{1}}\phi_{10})
 -t^-_{p_\pm p_z}(w^\prime+w \phi_{10})  \\ 
 -t^+_{p_\pm p_z}(w+w^\prime \phi_{10}) + t^-_{p_\pm p_z}(w^\prime\phi_{0\bar{1}}+w\phi_{0\bar{1}}\phi_{10})
 \emat,&
\eea 
and finally the coupling orbital terms in the $(\tau,p_\pm) \times (\kappa, s)$ sub-spaces

\bea 
&\hat{C}_{\kappa p_\pm}^{AA} = \bmat 0 && -t^-_{\kappa p_\pm}w && t^+_{\kappa p_\pm}w^\prime \\ 
0 && t^+_{\kappa p_\pm}w^\prime && -t^-_{\kappa p_\pm}w\emat \quad , \quad
\hat{C}_{\kappa p_\pm,1}^{AB} = \bmat t^+_{\kappa p_\pm}\phi_{0\bar{1}}\phi_{10} -t^-_{\kappa p_\pm}\phi_{10} &&
t^+_{\kappa p_\pm}w \phi_{10} && -t^-_{\kappa p_\pm}w^\prime \phi_{0\bar{1}} \phi_{10}\\ t^+_{\kappa p_\pm}\phi_{10}-t^-_{\kappa p_\pm}\phi_{0\bar{1}}\phi_{10} && -t^-_{\kappa p_\pm}w^\prime \phi_{10} && t^+_{\kappa p_\pm}w \phi_{0\bar{1}} \phi_{10} \emat,& \nonumber \\ \nonumber \\
&\hat{C}_{\kappa p_\pm,2}^{AB} = \bmat t^+_{\kappa p_\pm} -t^-_{\kappa p_\pm}\phi_{0\bar{1}} && t^+_{\kappa p_\pm}\phi_{0\bar{1}} -t^-_{\kappa p_\pm} \\
t^+_{\kappa p_\pm}w^\prime \phi_{0\bar{1}} && -t^-_{\kappa p_\pm}w \phi_{0\bar{1}} \\ 
-t^-_{\kappa p_\pm}w && t^+_{\kappa p_\pm}w^\prime \emat.& \\ \nonumber
\eea 
        
\end{widetext}
    

\section{boundary Green Function and Dyson Equations}\label{App_bGF}

To simplify notation, we omit the superscript `$A$' in advanced GFs from now on. Given the real-space components of the bulk GF in Eq.~\eqref{residues}, we next extend the method of Refs.~\cite{Liliana2009,Zazunov2016, Pablo2008} to derive 
the bGF characterizing a \emph{semi-infinite} 2D dimensional system. To that effect, we add
an impurity potential line $\epsilon$ localized at the frontier region \cite{Cristina2020}.  Taking the limit $\epsilon\to \infty$,
the infinite system is cut into disconnected semi-infinite subsystems with $j\leq-1$ (left side, $L$) and $j\geq1$ (right side, $R$). 
Using the Dyson equation, the local GF components of the cut subsystem follow as \cite{Zazunov2016}

\be \label{Dyson-n}
\hat{\mathcal{G}}_{j j} = \hat{G}^{(0)}_{j j} -\hat{G}^{(0)}_{j 0} \left[\hat{G}^{(0)}_{00}\right]^{-1} \hat{G}^{(0)}_{0 j},
\ee
where $\hat{G}^{(0)}$ are the bulk GF and $\hat{\mathcal{G}}$ are the semi-infinite GF. The bGF for the left and right semi-infinite system, respectively, are with Eq.~\eqref{Dyson-n} given by

\be
\hat{\mathcal{G}}_L(k_\parallel,\omega) = \hat{\cal G}_{\bar{1}\bar{1}}(k_\parallel,\omega)\quad ,\quad 
\hat{\mathcal{G}}_R (k_\parallel,\omega)=\hat{\cal G}_{11}(k_\parallel,\omega) .
\ee

Using this boundary Green function we can compute the spectral properties of open systems but also the spectral and transport properties for different type of junctions defined on MATBLG.

For instance, to determine the local spectral density at a domain wall \cite{Parameswaran2020, Ryu2020} where the coupling to the substrate changes sign, we use the following expression
%
\be
\hat{G}_{nn} = \hat{\mathcal{G}}_{nn} + \hat{\mathcal{G}}_{n1}\hat{\Sigma}^\dag_{LR}\left[ \hat{\mathbb{I}} - \hat{\mathcal{G}}_L\hat{\Sigma}_{LR}\hat{\mathcal{G}}_R\hat{\Sigma}^\dag_{LR} \right]^{-1} \hat{\mathcal{G}}_L \hat{\Sigma}_{LR} \hat{\mathcal{G}}_{1n}, 
\ee
where $\hat{\Sigma}^\dag_{LR}$ is the coupling term between the different regions that the define wall and requires the non-local GF for the right semi-infinite system given by

\bea 
&\hat{\mathcal{G}}_{n1} = \hat{G}^{(0)}_{n1} - \hat{\mathcal{G}}_R\hat{T}^\dag\hat{G}^{(0)}_{n1} \left[\hat{G}^{(0)}_{00} \right ]^{-1}\hat{G}^{(0)}_{01},& \nonumber \\
&\hat{\mathcal{G}}_{1n} = \hat{G}^{(0)}_{1n} - \hat{G}^{(0)}_{10} \left[\hat{G}^{(0)}_{00} \right ]^{-1} \hat{G}^{(0)}_{1n}\hat{T} \hat{\mathcal{G}}_R,&
\eea 
where $\hat{T}$ is the coupling term between consecutive sites of each semi-infinite system and

\bea 
&\hat{G}^{(0)}_{1n} = \hat{G}^{(0)}_{00}\left(\hat{T}\hat{\mathcal{G}}_R \right)^{n-1},& \nonumber \\
&\hat{G}^{(0)}_{n1} = \left(\hat{\mathcal{G}}_R \hat{T}^\dag \right)^{n-1}\hat{G}^{(0)}_{00}. &
\eea 
%
%
%
%
%
%
%
%
%

For computing transport properties in junctions we start from the general expression for the current between two leads using Keldysh Green functions in \cite{Cuevas1996,Zazunov2016}, which we adapt to the 1 valley MATBLG case

\bea
I_{LR}(k_\parallel) = \frac{2e}{h} \int d\omega \; &\textrm{tr}& \{\hat{\Sigma}_{LR} \hat{G}_{RR}^{-+} \hat{\Sigma}_{RL} \hat{\cal G}_L^{+-} \nonumber \\
&& - \hat{\Sigma}_{LR}\hat{G}_{RR}^{+-} \hat{\Sigma}_{RL} \hat{\cal G}_L^{-+} \},
\eea
where the factor $2$ comes from the spin degree of freedom. Finally we expand the non-equilibrium correlation functions in terms of advanced GF as using Langreth rules

\bea
\hat{G}^{+-/-+}_{RR} &=& (\hat{\mathbb{I}}+ \hat{G}_{LR}^\dag \hat{\Sigma}_{LR})\hat{\cal G}^{+-/-+}_R(\hat{\mathbb{I}}+\hat{\Sigma}_{LR}^\dag \hat{G}_{LR})+ \nonumber \\ &&\hat{G}^\dag_{RR} \hat{\Sigma}_{LR}^\dag \hat{\cal G}^{+-/-+}_L \hat{\Sigma}_{LR} \hat{G}_{RR}, 
\eea
with

\bea
&\hat{\cal G}^{+-}_j = f_j(\omega - \mu_j)(\hat{\cal G}_j - \hat{\cal G}_j^\dag),& \nonumber \\ 
&\hat{\cal G}^{+-}_j = \left(f_j(\omega - \mu_j)-1 \right)(\hat{\cal G}_j - \hat{\cal G}_j^\dag),&
\eea
$j=L/R$ and $f_j(\omega - \mu_j)$ is the Fermi-Dirac distribution. Using Dyson equation we can obtain the current in terms of the bGF

\bea 
&\hat{G}_{LR} = \hat{\cal G}_L\hat{\Sigma}_{LR} \hat{G}_{RR},& \nonumber \\ 
&\hat{G}_{RR} = [\hat{\mathbb{I}} -  \hat{\cal G}_R \hat{\Sigma}_{LR}^\dag \hat{\cal G}_L \hat{\Sigma}_{LR}]^{-1}\hat{\cal G}_R&.
\eea  

After some algebra we get the Landauer formula 

\be
I_{LR}(k_\parallel) = \frac{2e}{h} \int d\omega \; T(k_\parallel, \omega) \left(f_R-f_L\right),
\ee 
where

\begin{widetext}
\bea
&T(k_\parallel, \omega) = \textrm{tr}\left \{ \left (\hat{\mathbb{I}}+ \hat{G}_{LR}^\dag \hat{\Sigma}_{LR} \right) \left (\hat{\cal G}_R - \hat{\cal G}_R^\dag \right ) \left (\hat{\mathbb{I}}+ \hat{\Sigma}_{LR}^\dag \hat{G}_{LR} \right)\hat{\Sigma}_{LR}^\dag \left(\hat{\cal G}_L - \hat{\cal G}_L^\dag \right) \right \},& \nonumber \\ \nonumber \\
&\bra T(\omega) \ket = \displaystyle\frac{1}{n_k} \sum\limits_{k_\parallel} T(k_\parallel, \omega), &
\eea
\end{widetext}
where $T(k_\parallel, \omega)$ defines a momentum resolved transmission coefficient and $\bra T(\omega) \ket$ is the normalized total transmission function, which can be associated with a two-terminal zero temperature conductance through $G=G_0\bra T(0) \ket$ where $G_0 = 2e^2/h$ is the conductance quantum.

\section{Recursive GF method}\label{App_rGF}

In cases where a non-uniform potential is applied the bGF can be calculated recursively. 
We define the recursive the bGF at an dimensionless n-site as 

\bea 
&\left[ \hat{\mathcal{G}}^{rc}_R(n) \right]^{-1} = \omega \hat{\mathbb{I}} - \hat{\mathcal{H}}_0(k_\parallel) - \hat{\mathcal{H}}_b(n) - \Sigma_R(n),& \nonumber \\
&\left[ \hat{\mathcal{G}}^{rc}_L(n) \right]^{-1} = \omega \hat{\mathbb{I}} - \hat{\mathcal{H}}_0(k_\parallel) - \hat{\mathcal{H}}_b(n) - \Sigma_L(n),&
\eea 
where $\hat{\mathcal{H}}_0(k_\parallel)$ is the local contribution defined in a stripe and the recursive expression of the self energy takes the form

\bea 
\Sigma_R(n) &=& \hat{T}_{LR}\left[ \hat{\mathcal{G}}^{rc}_R(n-1) \right]^{-1}\hat{T}^\dag_{LR}, \nonumber \\
\Sigma_L(n) &=& \hat{T}^\dag_{LR}\left[ \hat{\mathcal{G}}^{rc}_L(n-1) \right]^{-1}\hat{T}_{LR}.
\eea 

As we can see the self-energy couples the n-site with the previous one and $n$ goes from $n=1$ to $n=N_x=L/2L_m$ where $L$ is the total length of the system. The self energy at the first site $\Sigma_{L/R}(1)$ can be defined to simulate the coupling to a doped normal lead with constant LDOS inducing a characteristic broadening of the states through the junction.

Inspired by the experimental device described in Ref.~\cite{Recher2020} the potential profile Hamiltonian for a three regions junction takes the form 

\bea 
&\hat{\mathcal{H}}_b(n) = \displaystyle\frac{U_b}{2}V_x(n)\hat{\mathbb{I}},& \nonumber \\
&V_x(n) = \tanh \left( \displaystyle\frac{n-(N_x-N_b)-1}{d_b} \right) +1.&
\eea 


For the 6B1V model and following the polynomial expansion in $z$ of the Hamiltonian $\hat{\mathcal{H}}(z)$ as described in \cite{forthcoming_paper} takes the form

\bea \label{FLA_poly}
\hat{\mathcal{H}}(z) &=& \sum_{i=1}^{2m+1} \hat{H}_i z^{i-(m+1)} \nonumber \\ &=& \hat{H}_1 z^{-m} +\dots+\hat{H}_{m+1}+ \dots + \hat{H}_{2m+1}z^m. \nonumber \\
\eea 
we can define the matrix that define the recursive method as

\be
\hat{\mathcal{H}}_0 = \bmat \hat{H}_2 && \hat{H}_3 \\
            \hat{H}_1 && \hat{H}_2 \emat \quad, \quad 
\hat{T}_{LR} = \bmat \hat{0} && \hat{H}_1 \\
        \hat{0} && \hat{0}\emat.
\ee

We note that this expression is totally general and can be use for any TB nearest neighbour Hamiltonian.


In the case of the 2B1V model, which goes beyond the nearest neighbors approximation, 
we follow Ref.~\cite{Recher2020} to set the recursive equations from 

\bea 
&\hat{\mathcal{H}}_0 = \bmat\hat{H}_0 && \hat{H}_1 && \hat{H}_2 && \hat{H}_3 \\
    \hat{H}_1^\dag && \hat{H}_0 && \hat{H}_1 && \hat{H}_2\\
    \hat{H}_2^\dag && \hat{H}_1^\dag && \hat{H}_0 && \hat{H}_1 \\
    \hat{H}_3^\dag && \hat{H}_2^\dag && \hat{H}_1^\dag && \hat{H}_0\emat,& \nonumber \\ \nonumber \\
&\hat{T}_{LR} = \bmat\hat{0} && \hat{H}_3 && \hat{H}_2 && \hat{H}_1 \\  
\hat{0} && \hat{0} && \hat{H}_3 && \hat{H}_2\\ 
\hat{0} && \hat{0} && \hat{0} && \hat{H}_3\\ 
\hat{0} && \hat{0} && \hat{0} && \hat{0}\emat,& 
\eea
where

\bea
\hat{H}_0 &=& \bmat 2t_2^\prime\sin{k_x L_x} && t_1e^{i\frac{k_x L_x}{2}}+t^*_4e^{-i\frac{k_x L_x}{2}} \\
t_1e^{-i\frac{k_x L_x}{2}}+t_4e^{i\frac{k_x L_x}{2}} && 2t_2^\prime\sin{k_x L_x} \emat, \nonumber \\ \nonumber \\
\hat{H}_1 &=& \bmat 0 && t_1+t_5^*e^{i k_x L_x} \\ t_1+t_5e^{-i k_x L_x} && 0 \emat, \nonumber \\ \nonumber \\
\hat{H}_2 &=& \bmat 0 && t_4^*e^{i\frac{k_x L_x}{2}}+t_5^*e^{-i\frac{k_x L_x}{2}} \\ t_4e^{-i\frac{k_x L_x}{2}}+t_5e^{i\frac{k_x L_x}{2}} && 0\emat, \nonumber \\ \nonumber \\
\hat{H}_3 &=& \bmat -2t_2^\prime\sin{\frac{k_x L_x}{2}} && t_5^* \\ t_5 && -2t_2^\prime\sin{\frac{k_x L_x}{2}}\emat.
\eea 

However, we equivalently can express the relevant matrices for the recursive method in terms of the polynomial expansion of the Hamiltonian.


\bibliographystyle{apsrev4-2}

\providecommand{\noopsort}[1]{}\providecommand{\singleletter}[1]{#1}%

\end{document}